\documentclass[aps,prl,twocolumn,showpacs,preprintnumbers, nofootinbib]{revtex4}

\usepackage{natbib}
\usepackage{graphicx}
\usepackage[english]{babel}
\usepackage{color}
\usepackage{amsmath,amssymb,amsfonts}
\usepackage{multirow}
\usepackage{slashed,ulem}

\newcommand{\be}{\begin{equation}}
\newcommand{\ee}{\end{equation}}
\newcommand{\bea}{\begin{eqnarray}}
\newcommand{\eea}{\end{eqnarray}}

\newcommand{\rmn}{\mathrm}

\begin{document}
\title{Is dark matter with long-range interactions \\a solution to all small-scale problems of $\Lambda$CDM cosmology?}

\author{Laura~G.~van~den~Aarssen}
\email{laura.van.den.aarssen@desy.de}
\affiliation{{II.} Institute for Theoretical Physics, University of Hamburg, 
Luruper Chausse 149, DE-22761 Hamburg, Germany}

\author{Torsten~Bringmann}
\email{torsten.bringmann@desy.de}
\affiliation{{II.} Institute for Theoretical Physics, University of Hamburg, 
Luruper Chausse 149, DE-22761 Hamburg, Germany}

\author{Christoph~Pfrommer}
\email{christoph.pfrommer@h-its.org}
\affiliation{Heidelberg Institute for Theoretical Studies, Schloss-Wolfsbrunnenweg 35, D-69118 Heidelberg, Germany}

\date{6 December 2012}

\begin{abstract}
The cold dark matter (DM) paradigm describes the large-scale structure of the universe remarkably well. However, there exists some tension with the observed abundances and internal density structures of both field dwarf galaxies and galactic satellites. Here, we demonstrate that a simple class of DM models may offer a viable solution to all of these problems simultaneously. Their key phenomenological properties are velocity-dependent self-interactions mediated by a light vector messenger and thermal production with much later kinetic decoupling than in the standard case.\end{abstract}

\pacs{95.35.+d, 95.85.Pw, 98.65.-r, 98.70.Sa}

\maketitle


\paragraph{Introduction.---}
Recent advances of cosmological precision tests further consolidate the
`cosmological concordance model',
indicating that 4.5\% of
the mass in the universe is in baryons, 22.6\% is non-baryonic cold dark matter 
(CDM), and the
rest is Einstein's cosmological constant $\Lambda$ (or behaves like it)
\cite{Komatsu:2010fb}.  The leading CDM candidates are weakly interacting
massive particles (WIMPs) which are thermally produced in the early universe
\cite{DM_reviews}. While their {\it chemical} decoupling from the heat bath sets
the observed DM relic density today, their {\it kinetic} decoupling
induces a small-scale cutoff in the primordial power spectrum of density
perturbations \cite{Green:2005fa}. For neutralino DM, e.g., this cutoff
corresponds to a smallest protohalo mass of $M_{\rm
  cut}/M_\odot\sim10^{-11}$--$10^{-3}$ \cite{Bringmann:2009vf}, but
  it could be as large as $M_{\rm cut}\gtrsim10\,M_\odot$ if DM couples to new light scalars \cite{vandenAarssen:2012ag}.  
  After kinetic decoupling, standard WIMP CDM behaves like a collision-less gas. Baryons, on the other hand,
can radiate away excess energy and sink to the centers of CDM halos where they
form stars and galaxies.  In this picture, structure formation proceeds
hierarchically with galaxies to form at sites of constructive interference of
small-scale waves in the primordial density fluctuations.

Despite the great success of $\Lambda$CDM cosmology, detailed observations of
nearby small galaxies pose a number of puzzles to this paradigm. Here, we
isolate three distinct classes of problems.  (1) The observed galaxy luminosity
and H{\sc i}-mass functions show much shallower faint-end slopes than predicted
by $\Lambda$CDM models \cite{Zavala+2009}; this is locally known as the `missing
satellites problem' of the Milky Way (MW), which should contain many more
dwarf-sized subhalos than observed \cite{Kravtsov2010}. (2) Simulations predict
an inner DM {\it cusp} for the density structure of galaxies, seemingly at odds
with the {\it cored} profiles found in observed low surface brightness galaxies
and dwarf satellites \cite{cuspcore}.  (3) Recently, it was realized that the
  most massive subhalos in $\Lambda$CDM simulations of MW-size halos have an
  internal density structure that is too concentrated in comparison to the
  observed brightest MW satellites: the simulated circular velocity profiles
  increase more steeply and attain their maximum circular velocity at smaller
  radii than any of the observed ones.  On the other hand, those simulated
  subhalos should be `too big to fail' in forming stars according to our
  understanding of galaxy formation (being more massive than the
  UV-photosuppression scale at all redshifts, after formation, for conceivable
  reionization histories). Thus, it is extremely puzzling why there is no observed
  analogue to those objects \cite{tbtfproblem}.

Astrophysical solutions to (1) invoke suppressing the formation of galaxies within existing dwarf halos or suppressing  the star formation in dwarf galaxies. Galaxy formation can be 
held back by increasing the gas entropy before collapse,
 e.g.~via photoionization \cite{photoheating}, blazar heating \cite{PCB} or AGN feedback in 
 the radio-quiet mode \cite{Silk:2010aw}. 
A photoionization-induced lack of H{\sc i} \cite{Haim-Rees-Loeb:97} or
intrinsically low metallicities \cite{Tass-Krav-Gned:08} may further
suppress the cooling efficiency of collapsing baryons. Numerical
simulations with a photoionizing background, however, cannot suppress dwarf galaxy
formation at the level implied by observations
\cite{Hoeft+2006}. In principle, gas may also be removed from 
dwarfs via photo-evaporation \cite{photoevap} and
feedback from supernovae
\cite{SNfeedback}.  Any such feedback, however, 
implies  remnant stellar populations and H{\sc i} masses 
in conflict with  most recent observational constraints \cite{Zwaan+2010}.

The `cusp-core' problem (2) may be addressed by large velocity
anisotropies or reduced central DM densities. There is a degeneracy
  between cored isotropic and cuspy anisotropic velocity distributions
  and the stellar line-of-sight velocity data is still too sparse to
  dynamically resolve (2) \cite{Breddels+2012}. Reducing central DM
  densities was proposed as a result of efficient baryonic feedback
  processes \cite{Pontzen+2011}, however in contradiction to cuspy
  dwarf profiles in other simulations with feedback
  \cite{Sawala+2010}.

The `too big to fail' problem (3) might be solved by either an
increased stochasticity of galaxy formation on these scales or a total MW mass
$\lesssim8\times10^{11}
\,\rmn{M}_\odot$ \cite{TBTFsolve}.
Abundance matching 
of stellar and halo mass, which agree with stacking analyses of gravitational
lensing signals and satellite dynamics of SDSS galaxies, make the required large degree of
stochasticity implausible  \cite{Guo+2010}. For a $10^{12} \,\rmn{M}_\odot$ MW, on the other hand,  
the chance to host two  satellites as massive as the Magellanic Clouds 
is less than 10\% \cite{Busha+2011} and even
lower  for smaller MW masses (from satellites studies of
MW-type SDSS systems \cite{SDSS} and MW and Andromeda orbit timing arguments \cite{LiWhite2008}).

The next logical possibility that could lead to  a suppression of small-scale power is a
modification of the CDM paradigm itself. The most often discussed options are 
interacting DM (IDM) \cite{Sper-Stei:00} and warm DM (WDM)
\cite{wdm}, though it should be noted that there exist interesting alternatives such as DM from late decays \cite{DM_decay}, DM with  large annihilation rates \cite{Kaplinghat:2000vt}, extremely light DM particles forming a condensate \cite{Hu:2000ke}, or inflationary models with broken-scale invariance \cite{inflation}.
 As was soon realized, however,
 IDM with a constant cross section produces spherical cores  in
conflict with observed ellipticities in clusters
\cite{cluster_ellipt} and the survivability of satellite halos
\cite{Gnedin+2001}. While WDM is unlikely to account for some of the large $\sim$1\,kpc cores
 claimed in dwarfs \cite{Maccio+2012}, 
 and severely constrained by Lyman-$\alpha$ observations \cite{Viel2006,Seljak2006,Boyarsky-WDM2009,Boyarsky+2009},
  it may be able to partially
resolve the `too big too fail' problem by allowing these subhalos to initially
form with lower concentrations \cite{Lovell+2011}. 
Alternatively,
 DM self-interactions mediated by a Yukawa potential, with the resulting
characteristic velocity dependence of the transfer cross section 
\cite{Feng:2009hw,Buckley:2009in}, avoid 
constraints on scales of MW-type galaxies and beyond \cite{Loeb:2010gj} and
produce $\sim$1\,kpc cores that match the observed
velocity profiles of massive MW satellites \cite{Vogelsberger:2012ku} (see also Ref.~\cite{Feng:2009hw}).

\begin{figure}[t]
	\includegraphics[width=0.98\columnwidth]{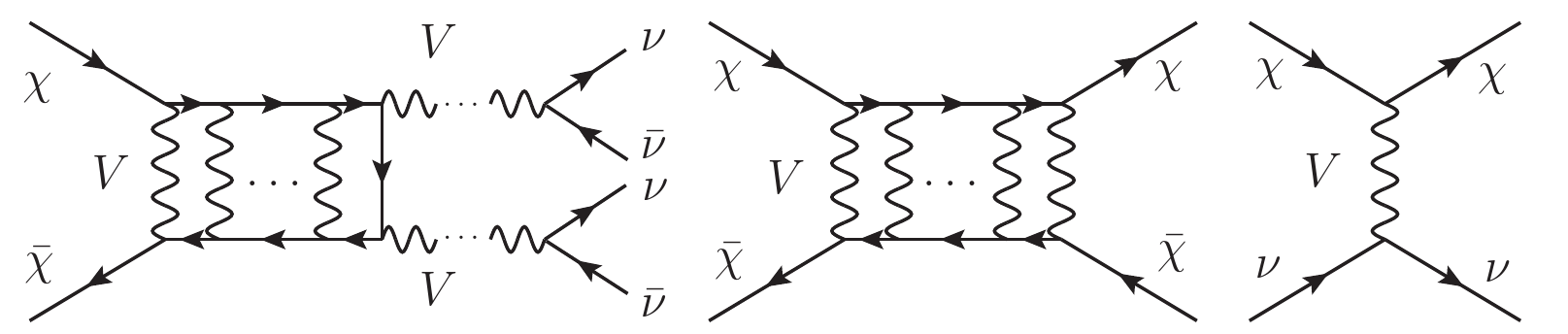}
	 \caption{Interaction processes that set the DM relic density and may lead to observable neutrino annihilation products today (left), change the inner velocity and density profile of dwarf halos (middle) and induce a comparatively large cutoff in the spectrum of primordial density perturbations (right).
	 	 } \label{fig:feynman}
\end{figure}

Most astrophysical and DM solutions have shortcomings, or can explain at most two
of the three problems, which makes them less attractive on the basis of Occam's razor. Here, we
demonstrate that there is a class of IDM models that {\it simultaneously} can
account for all three problems. Encouraged by the results of
Refs. \cite{Loeb:2010gj,Vogelsberger:2012ku}, in particular, we will focus on
models with a Yukawa-like interaction between the DM particles that is mediated
by a light messenger (see Fig.~\ref{fig:feynman}).  As we will show, the kinetic 
decoupling of DM in these
models can happen sufficiently late to suppress the power spectrum at scales 
as large as that of dwarf galaxies,
  $M_{\rm cut}\gtrsim 10^9M_\odot$,
while at the same time the velocity-dependent self-interaction of DM produces
cored density profiles in dwarfs \footnote{Late kinetic decoupling ($T_{\rm
    kd}\sim0.1\,$keV) was advocated as potential solution to the missing
  satellites problem before \cite{Boehm:2000gq} -- though this analysis
  considerably under-estimated $T_{\rm kd}$ for WIMPs \cite{Chen:2001jz}.  }.

\paragraph{Model setup.---}
In models with new light exchange particles $\phi$,  kinetic decoupling  can  happen much later than in standard WIMP scenarios, in particular for small masses  $m_\phi$ \cite{vandenAarssen:2012ag}. For {\it scalar} exchange particles, however, the amplitude for DM scattering with leptons scales like $\sim m_\chi m_\ell /m_\phi^2$, implying that scattering with neutrinos is generally negligible.
While a coupling of $\phi$ to charged leptons also leads to a loop-suppressed  effective coupling to photons, $\mathcal{L}\supset g_{\phi\gamma\gamma}\phi F^{\mu\nu}F_{\mu\nu}$, the resulting scattering amplitude does not contribute in the relevant limit of small momentum transfer.
Kinetic decoupling therefore never occurs at $T_{\rm kd}\ll0.1\,$MeV, at which point the number density of electrons starts to become strongly Boltzmann-suppressed
and there are no lighter (and thus more abundant) particle species left that could keep up kinetic 
equilibrium instead.  

Let us consider instead  the situation where DM consists of heavy Dirac fermions $\chi$ which only couple
to a light {\it vector} boson $V$. Due to our interest in late kinetic decoupling, we will require $V$ to also couple to neutrinos:
\be
  \label{Lint}
  \mathcal{L}_{\rm int} \supset -g_\chi \bar\chi \slashed V \chi - g_\nu \bar\nu \slashed V \nu\,.
\ee
Note that we take a phenomenological approach here and only state couplings 
 that explicitly enter our analysis. In particular, $V$ does not have to be a gauge 
boson, which leaves  couplings to other SM particles unspecified (see 
e.g.~Ref.~\cite{Chiang:2012ww}  for a recent model-independent analysis).
 DM is then thermally produced in the early universe via $\bar \chi\chi\leftrightarrow VV$.
Assuming  $g_\nu$ is small, but large enough to thermalize $V$ at early times, the  relic density  is   given by
\be
 \label{omega}
  \Omega_\chi h^2= \Omega_{\bar\chi} h^2\simeq \frac{0.11}{2}\left(\frac{g_\chi}{0.683}\right)^{-4}\left(\frac{m_\chi}{\mathrm{TeV}}\right)^{2}\,.
\ee
This expression receives $\mathcal{O}(1)$ corrections due to the Sommerfeld effect \cite{sommerfeld}, i.e.~a multiple exchange of $V$ as shown in Fig.~\ref{fig:feynman}, which we fully take into account in our analysis.
The kinetic decoupling temperature, on the other hand,  will be set by $\chi$-$\nu$ scattering. The corresponding  amplitude at small momentum transfer reads 
\be
  \label{scatteringM2}
   \sum_{\rm all~spins}\left|\mathcal{M}\right|^2_{\chi \nu\leftrightarrow \chi \nu}= 64 g_\chi^2g_\nu^2 \frac{m_\chi^2E_\nu^2}{m_{V}^4}\,.
\ee

In the following, we will consider $g_\nu$ as an essentially free parameter while 
$g_\chi$ is fixed by the requirement to obtain the correct  relic density (see e.g.~Ref.~\cite{Fox:2008kb} for a list of possible natural explanations for $g_\nu\ll g_\chi$).

\paragraph{DM self-scattering.---}
The light vector messenger induces a long-range attractive Yukawa potential between the DM
particles, cf.~Fig.~\ref{fig:feynman}. Concerning elastic DM self-scattering, this
 is  completely analogous to screened Coulomb scattering in a
plasma for which simple parametrizations of the transfer cross
section $\sigma_T(v)$ in terms of $m_\chi$, $m_{V}$, $g_\chi$ and the relative 
velocity $v$ of the DM particles exist
\cite{yukawa_param,Feng:2009hw}. Using these
parametrizations, it was shown that the type of DM model introduced above
produces cores rather than cusps \cite{Loeb:2010gj} and may solve the `too big too fail problem' 
\cite{Vogelsberger:2012ku}, without being in conflict with the strong
constraints for models with constant $\sigma_T$. We also note that
$\sigma_T$ drops with larger $v$ such that  for galaxy {\it clusters} only the very central density profile  at $r\lesssim\mathcal{O}(1-10)\,$kpc  will be smoothed  out, matching
observational evidence (from improved lensing
and stellar kinematic data \cite{Newman:2011}) for a density cusp in A383  that is slightly shallower than expected for standard CDM.

For our discussion, the astrophysically important quantities are the velocity $v_{\rm max}^2=g_\chi^2m_{V}/(2\pi^2m_\chi)$ at which $\sigma_T v$ becomes maximal and $\sigma_T^{\rm max}\equiv\sigma_T(v_{\rm max})=22.7\,m_{V}^{-2}$. In particular, $v_{\rm max}$ should not be too different from the typical velocity dispersion $\sigma_v\sim\mathcal{O}(10)\,$km/s encountered in dwarf galaxies if one wants to make any contact to potential problems with standard structure formation at these scales. On the other hand, the value of $\sigma_T^{\rm max}$ is constrained by various astrophysical measurements, see Ref.~\cite{Vogelsberger:2012ku} for a compilation of current bounds. 

\begin{figure}[t]
	\includegraphics[width=0.9\columnwidth]{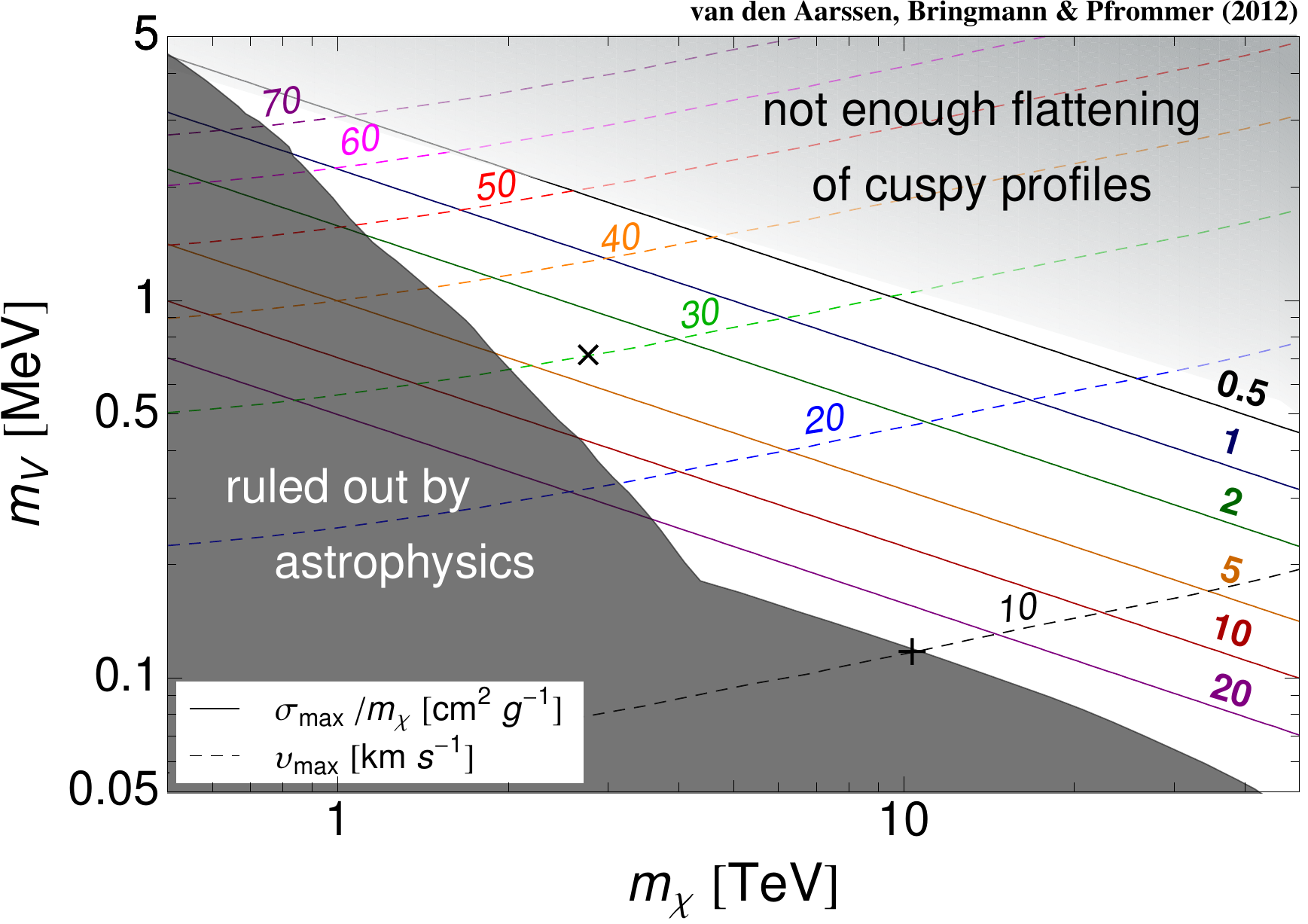}
         \vspace*{-0.15cm}
	 \caption{ 
	 The white area corresponds to  DM and mediator  masses that  may
	 solve the `cusp vs.~core' problem.  
	 The crosses indicate two benchmark models for which detailed simulations  \cite{Vogelsberger:2012ku} have found a solution to the `too big to fail' problem. Dashed and solid lines show contours of the astrophysical relevant quantities $\sigma^T_{\rm max}$ and $v_{\rm max}$. See text for further details.
	 	 } \label{fig:self_scatter}
\end{figure}

Fixing $g_\chi$ by the relic density requirement, there is a one-to-one correspondence between the particle physics input $(m_\chi,m_{V})$ and the astrophysically relevant parameters $(v_{\rm max},\sigma_T^{\rm max})$. As demonstrated  in Fig.~\ref{fig:self_scatter}, a solution to the aforementioned small-scale problems (2) and (3) may then indeed be possible for DM masses of $m_\chi\gtrsim600\,$GeV and a mediator mass  in the (sub-) MeV range. We also display the strongest astrophysical bounds on large DM self-interaction rates \cite{Loeb:2010gj}. For $m_\chi\lesssim4\,$TeV, they arise from collisions with particles from the dwarf parent halo, while at larger $m_\chi$  an imminent gravothermal catastrophe  is more constraining.

\begin{figure}[t]
	\includegraphics[width=0.9\columnwidth]{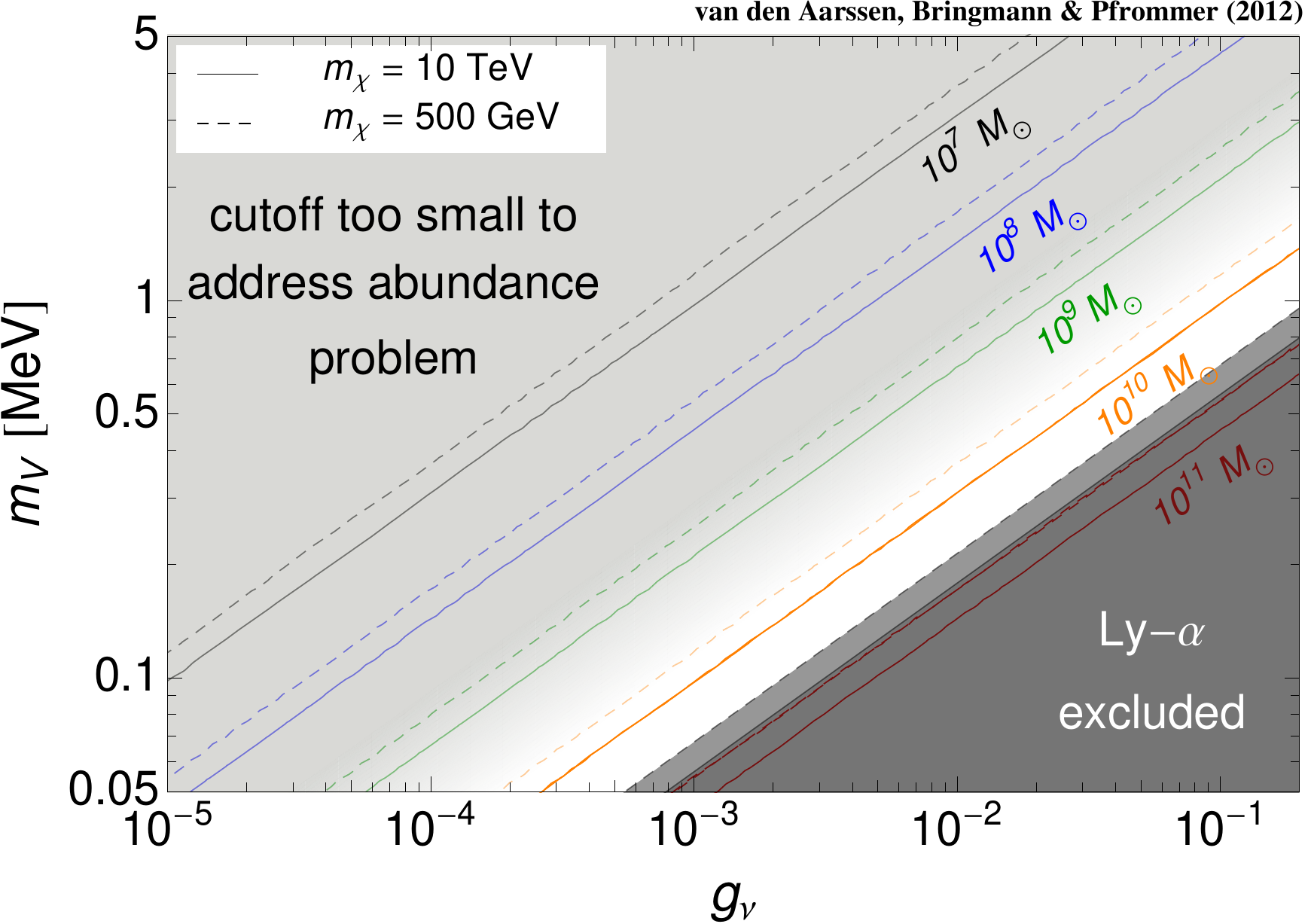}
         \vspace*{-0.15cm}
	 \caption{This plane shows the mediator mass $m_{V}$ vs.~the coupling strength $g_\nu$. Large values of $g_\nu$ and small values of $m_{V}$ lead to late kinetic decoupling and thus a large mass $M_{\rm cut}$ of the smallest protohalos. $M_{\rm cut}\gtrsim5\times10^{10}M_\odot$ is excluded by Ly-$\alpha$ data while $M_{\rm cut}\gtrsim10^{9}M_\odot$ may solve the small-scale abundance problems of $\Lambda$CDM cosmology.} \label{fig:mg}
\end{figure}

\paragraph{The small-scale cutoff.---}
For small kinetic decoupling temperatures $T_{\rm kd}$, acoustic oscillations \cite{ao} are more efficient than free streaming effects to suppress the power spectrum \cite{Bringmann:2009vf,{1205.1914}}. The resulting exponential cutoff can be translated into a smallest protohalo mass  of
\be
   \label{mcut}
   M_{\rm cut}\approx \frac{4\pi}{3}\left.\frac{\rho_\chi}{H^3}\right|_{T=T_{\rm kd}}= 1.7\times10^8\left(\frac{T_{\rm kd}}{\rm keV}\right)^{-3} M_\odot\,,
\ee
where $H$ is the Hubble rate and we assumed 
late kinetic decoupling such that the 
 effective number of relativistic degrees of freedom $g_{\rm eff}=3.37$.
For scattering with relativistic neutrinos, c.f.~Eq.~(\ref{scatteringM2}), the analytic treatment of kinetic decoupling given in Ref.~\cite{Bringmann:2006mu} is valid. Extending those expressions to allow for $T_\nu\neq T$, we find
\be
  \label{tkd}
 T_{\rm kd} = 
  \frac{0.062\,{\rm keV}}{N_\nu^\frac14 \left({g_\chi}
   {g_\nu}\right)^\frac12} 
    \left(\frac{T}{T_\nu}\right)^\frac12_{\rm kd}
    \left(\frac{m_\chi}{\rm TeV}\right)^\frac14
    \left(\frac{m_{V}}{\rm MeV}\right)\,,
\ee
where $N_\nu$ is the number of neutrino species coupling to $V$.
Combining this with Eq.~(\ref{omega}) we therefore expect that $T_{\rm kd}$, and thus $M_{\rm cut}$, is essentially independent of  $g_\chi$ and $m_\chi$.

Using for definiteness $N_\nu=3$ and $T_\nu=(4/11)^\frac13T_\gamma$, we show in Fig.~\ref{fig:mg} contours of constant $M_{\rm cut}$ in the  ($g_\nu$,$m_{V}$) plane. We find that the result of the full numerical calculation \cite{Bringmann:2009vf,vandenAarssen:2012ag} is indeed extremely well described by Eqs.~(\ref{mcut},\ref{tkd}) for $g_\nu\gtrsim10^{-7}$ (assuming $m_\chi\sim1\,$TeV and $m_{V}\sim1\,$MeV; this value is even lower for larger $m_\chi$ and smaller $m_{V}$). For $g_\nu\lesssim10^{-7}$, DM scattering with the non-relativistic mediator particles $V$  starts to dominate over scattering with neutrinos, and $M_{\rm cut}$ eventually becomes independent of $g_\nu$. We checked that the new era of DM annihilation generally expected in models with Sommerfeld-enhanced annihilation rates  (see Ref.~\cite{vandenAarssen:2012ag} for a consistent treatment) has only negligible impact on our results for the late decoupling times  we focus on here.

\paragraph{Lyman-$\alpha$ forest bounds.---}

Conventionally, a possible cutoff in the power spectrum is often expressed in terms
of the mass $m_{\rm w}$ of a WDM thermal relic. In this case, it is set by free streaming
of the WDM particles and the comoving free-streaming
length $R_{\rmn{f}}$ is given by 
$R_{\rmn{f}} = 0.1 \,\left({\Omega_{\rmn{m}}h^2}/{0.13}\right)^{1/3}\,
  \left({m_{\rmn{w}}}/{\rmn{keV}}\right)^{-4/3}\,\rmn{Mpc}$
\cite{Sommer-Larsen2001}.
For a characteristic wavenumber $k_{\rmn{f}}
\equiv 0.46/R_{\rmn{f}}$,  the linear perturbation amplitude is
suppressed by a factor of 2  and  the characteristic
filtering mass can be defined as \cite{Sommer-Larsen2001}
\begin{equation}
  \label{eq:Mf}
   M_{\rmn{f}} \equiv \frac{4 \pi}{3}\,\bar{\rho}_{\rmn{m}}\,
   \left(\frac{\lambda_{\rmn{f}}}{2}\right)^3
   = 5.1\times 10^{10} \left(\frac{m_{\rmn{w}}}{\rmn{keV}}\right)^{-4}\rmn{M}_\odot\,,
\end{equation}
where $\lambda_{\rmn{f}} = 2\pi/k_{\rmn{f}}\simeq 13.6 R_{\rmn{f}}$. 
This choice of $M_{\rmn{f}}$ is justified by numerical experiments
\cite{Tittley1999} that find the resulting halo statistics for an
initial density distribution with a sharp cut in the power spectrum at
$k_{\rmn{c}}=2\pi/\lambda_{\rmn{f}}$ to be very similar to the
statistics of an initial density field smoothed with a top-hat window
of radius  $\lambda_{\rmn{f}}/2$.
Cosmological WDM simulations show a deviation
of the mass function from the CDM case on scales given by Eq.~\eqref{eq:Mf}
\cite{Avila-Reese2001,Zavala+2009}.

Combining data of the Lyman-$\alpha$ forest, the cosmic microwave background
(CMB) and galaxy clustering allows to constrain the cutoff scale in the
power spectrum; in terms of the mass of a thermal WDM candidate, a $2\sigma$-bound of
 $m_{\rmn{w}}>2\,$keV has been claimed 
\cite{Viel2006,Seljak2006}. This  weakens to $m_{\rmn{w}}>0.9\,$keV
when rejecting less reliable data at $z>3.2$ \cite{Viel2006} due to
systematic errors \cite{Boyarsky-WDM2009}. Revisiting
Lyman-$\alpha$ data yielded $m_{\rmn{w}}>1.7\,$keV 
which, however, is subject to systematic uncertainties at the $\sim$30\%
level \cite{Boyarsky+2009}, especially considering that blazar heating was not accounted for
 in deriving cosmological constraints
\cite{Puchwein+2011}. 

Lyman-$\alpha$ data thus firmly exclude
  $m_{\rmn{w}}<1\,$keV or $M_{\rmn{f}} > 5.1\times
10^{10}\,\rmn{M}_\odot$ (corresponding to a maximal circular velocity
$v_\rmn{max}\sim70\,\rmn{km~s}^{-1}$).  For $m_\rmn{w}\simeq 1-2$~keV, WDM models
are able to alleviate the `missing satellite problem' (somewhat depending on
feedback recipes) \cite{Maccio_WDM2010}, bring the faint-end of the galaxy
luminosity function into agreement with data \cite{Menci+2012}, and
match \cite{Zavala+2009} the H{\sc i} velocity function measured in the ALFALFA survey
\footnote{The small-end of simulated velocity
  functions for a thermal relic WDM particle with $m_\rmn{w}\lesssim 3$~keV
  are noticeably shallower than the analytical estimates of the Sheth and Tormen
  \cite{Sheth-Tormen1999}.}. For $m_\rmn{w}> 3$~keV, the corresponding mass
cutoff $M_{\rmn{f}} < 6\times 10^{8}\,\rmn{M}_\odot$ is too small to have any
impact on the faint-end of the galaxy luminosity function.
We include these bounds in Fig.~\ref{fig:mg} to demonstrate that our model
can successfully address  also the abundance problem (1).

\paragraph{Discussion.---}

In a phenomenological approach to identify the key properties of DM models that 
can  address all three $\Lambda$CDM small-scale problems  simultaneously, 
we found it sufficient to simply postulate the existence of a light vector messenger $V$ 
that couples to both DM and neutrinos as  in Eq.~(\ref{Lint}). 
If $V$ does not couple to quarks or other leptons, the coupling 
  $g_\nu$ is essentially unconstrained 
\cite{Chiang:2012ww}. While beyond the scope of this letter, however, we stress
that it would be very worthwhile to study possible concrete realizations of our  setup. 

The  greatest challenge for such a model building might be to prevent,
even at the one-loop level,  a kinetic mixing between $V$ and photons, which is severely 
constrained  for $m_{V}\lesssim$ MeV \cite{Jaeckel:2010ni}. 
On the other hand, limits on tree-level couplings of $V$ to charged leptons (e.g.~from
contributions to the anomalous magnetic moment of the
muon, beam dump experiments or low-$|q|^2$ $\nu$-$e$ 
scattering \cite{Jaeckel:2010ni,morelimits}) seem less severe and could at least partially be evaded 
by generation-specific  couplings.
Another option could be a new $U(1)$ coupling to DM and {\it sterile} neutrinos $\nu_s$ 
\cite{Harnik:2012ni}. As long as the $\nu_s$ have 
been in equilibrium in the very early universe and are  relativistic at 
$T_{\rm kd}$, this would not change the phenomenology of our model; Eq.~(\ref{tkd}),
in particular, would still apply.
It is therefore quite interesting that  CMB observations seem to favor 
additional relativistic degrees of freedom, which corresponds to the presence of  one light 
sterile neutrino species \cite{Archidiacono:2011gq}.

Finally, we note that for typical galactic  velocities $v\sim10^{-3}$, the type of DM candidate we propose here  annihilates with a Sommerfeld-enhanced rate of $\langle\sigma v\rangle\sim 3\times 10^{-24}$ $(m_\chi/{\rm TeV})^{-2}$cm$^3$s$^{-1}$ into a  $VV$ pair  which then decays {\it exclusively} into neutrinos (if $m_{V}\leq2m_e$). 
Such a large annihilation rate will be in reach of future IceCube observations 
of the galactic center \cite{IceCube}; for $m_\chi\lesssim1\,$TeV, in fact, a strong   Sommerfeld-induced  substructure enhancement of the signal \cite{Bovy:2009zs} may already be constrained.

\paragraph{Conclusions.---}
We have introduced a class of DM models with the unique property of addressing all three $\Lambda$CDM small-scale problems {\it simultaneously}, which should make them very attractive alternatives to be studied. From a model-building point of view, the only ingredient that is needed is a (sub-)MeV vector messenger particle that weakly couples to neutrinos and even more weakly to other standard model particles. While collider and direct searches for DM will be extremely challenging in this scenario, a TeV neutrino signal from the galactic center could turn out to be a smoking gun signature.

{\it Acknowledgements.} 
We thank J\"orn Kersten, Alessandro Mirizzi, Thomas Schwetz, Joe Silk, Mark Vogelsberger, Simon White and Jesus Zavala for very useful communications and feedback on the manuscript.
L.v.d.A. and T.B. acknowledge support from the German Research Foundation (DFG) through the Emmy Noether  grant BR 3954/1-1. C.P. gratefully acknowledges financial support of the Klaus Tschira Foundation.

\end{document}